# Automated Segmentation and Volume Measurement of Intracranial Carotid Artery Calcification on Non-Contrast CT

Gerda Bortsova, MSc,[a] Daniel Bos, MD, PhD,[b] Florian Dubost, PhD,[c] Meike W. Vernooij, MD, PhD,[b,d] M. Kamran Ikram, MD, PhD,[b] Gijs van Tulder, MSc,[e] Marleen de Bruijne, PhD[a,f]

From the [a]Biomedical Imaging Group Rotterdam, Department of Radiology and Nuclear Medicine, Erasmus MC, Rotterdam, The Netherlands; [b]Department of Epidemiology, Erasmus MC, Rotterdam, The Netherlands; [c]Department of Biomedical Data Science, Stanford University, The United States of America; [d]Department of Radiology and Nuclear Medicine, Erasmus MC, Rotterdam, The Netherlands; [e]Faculty of Science, Radboud University, The Netherlands; [f]Machine Learning Section, Department of Computer Science, University of Copenhagen, Copenhagen, Denmark.

The current study originated from Erasmus Medical Center (MC), Doctor Molewaterplein 40, 3015 GD Rotterdam, The Netherlands. Correspondence to Gerda Bortsova, MSc, Erasmus MC, P.O. Box 2040, 3000 CA Rotterdam, The Netherlands. E-mail: g.bortsova@erasmusmc.nl; tel.: +31-10-7038875; fax: +31-10-7034033.

This research is part of the research project Deep Learning for Medical Image Analysis (DLMedIA), project number P15-26, funded the Dutch Technology Foundation STW, which is part of the Netherlands Organisation for Scientific Research (NWO), and which is partly funded by the Ministry of Economic Affairs.

**This manuscript has been accepted for publication in Radiology: Artificial Intelligence (https://pubs.rsna.org/journal/ai), which is published by the Radiological Society of North America (RSNA), and has received the following DOI: 10.1148/ryai.2021200226.**

**Summary:**

**Automated deep-learning-based segmentation and volume measurement of intracranial carotid artery calcification (ICAC) had similar accuracy to manual assessment by trained observers; automated ICAC volume was also associated with incident stroke.**

**Key points**:

- Automated delineation of intracranial calcifications reached sensitivity of 83.8% and positive predictive value (PPV) of 88.0%; this performance compared well to that of manual assessment (73.9% sensitivity and 89.5% PPV).
- In visual examinations of 294 regions by a blinded expert, automated delineations were judged as more accurate than manual delineations in 131 regions, less accurate in 94 regions, and equally accurate in the rest of the regions (131 of 225; P = .01).
- Both automated and manual calcium volumes were associated with incident stroke, with adjusted hazard ratios of 1.38 (95% CI: 1.12, 1.75) and 1.48 (95% CI: 1.20, 1.87), respectively.

**Abbreviations**:

FPV = false positive volume, ICAC = intracranial internal carotid artery calcification, SD = standard deviation

# Abstract

## Purpose

To develop and evaluate a fully-automated deep-learning-based method for assessment of intracranial carotid artery calcification (ICAC).

## Methods

This was a secondary analysis of prospectively collected data from the Rotterdam study (2003-2006) to develop and validate a deep-learning-based method for automated ICAC delineation and volume measurement. Two observers manually delineated ICAC in non-contrast CT scans of 2,319 participants (mean age 69 ± 7 years; 1154 women [53.2%]) and a deep learning model was trained to segment ICAC and quantify its volume. Model performance was assessed by comparing manual and automated segmentations and volume measurements to those produced by an independent observer (available in 47 scans), comparing the segmentation accuracy in a blinded qualitative visual comparison by an expert observer, and comparing the association with first stroke incidence from the scan date until 2016. All method performance metrics were computed using 10-fold cross-validation.

## Results

The automated delineation of ICAC reached sensitivity of 83.8% and positive predictive value (PPV) of 88%. The intraclass correlation between automatic and manual ICAC volume measures was 0.98 (95% CI: 0.97, 0.98; computed in the entire dataset). Measured between the assessments of independent observers, sensitivity was 73.9%, PPV was 89.5%, and intraclass correlation was 0.91 (95% CI: 0.84, 0.95; computed in the 47-scan subset). In the blinded visual comparisons of 294 regions, automated delineations were judged as more accurate than manual delineations in 131 regions, less accurate in 94 regions, and equally accurate in the rest of the regions (131 of 225; 58.2%; P = .01). The association of ICAC volume with incident stroke was similarly strong for both automated (hazard ratio, 1.38 [95% CI: 1.12, 1.75] and manually measured volumes (hazard ratio, 1.48 [95% CI: 1.20, 1.87]).

## Conclusions

The developed model was capable of automated segmentation and volume quantification of ICAC with accuracy comparable to human experts.

# Introduction

Intracranial arteriosclerosis is a major risk factor for stroke (1, 2) and has been linked to an increased risk of dementia (3). An important indicator of intracranial arteriosclerosis is intracranial internal carotid artery calcification (ICAC), which can be visualized using CT (4-5). ICAC is thus a promising imaging marker for the assessing risk of cerebrovascular diseases. However, the quantitative measurements of ICAC currently rely on time-consuming and error-prone manual annotations (6, 7-8). Automating ICAC assessment could therefore facilitate research on the etiology and clinical consequences of intracranial arteriosclerosis and may ultimately enable the use of ICAC assessment in clinical practice.

Automating ICAC detection on CT is challenging due to its proximity to bony structures with similar attenuation, similarity in appearance to other structures (eg dural calcifications), and image artifacts. These challenges complicate the application of simple image processing techniques. Machine learning methods, however, are very suitable for this task as they can "learn" from examples without being explicitly programmed. Machine learning (including deep learning (9, 10)) has been extensively applied to assessing arterial calcification in vessel beds other than the intracranial carotid arteries, including the coronary arteries, aorta, cardiac valves, and extracranial carotid arteries (11-14). To date, only our previous study (15) has proposed a method for automating ICAC assessment; however, in this study there was no comparison of model performance with intra- or inter-observer agreement or variability, no visual assessment of the quality of the automated segmentations, nor were any associations determined between model outputs and clinically relevant factors.

The purpose of this study was to evaluate deep-learning-based automated ICAC assessment in terms of its accuracy and clinical value for stroke risk estimation. In the process of validating our model, we assessed our model in three ways: *(a)* comparison of model segmentation and volume measurement accuracy to an independent observer, using another observers annotations as the references standard, *(b)* comparison of the accuracy of automated segmentations to that of manual segmentations in a qualitative, visual manner (with analysis of error patterns in both manual and automated segmentations); and *(c)* assessment of ICAC presence and volume in association with stroke for automated and manual ICAC volume measures.

## Materials and Methods

### Study Participants

The current study focused on a sample of 2,319 participants (mean age 69 years, aged 55 or older; 1154 women [53.2%]) from the Rotterdam Study (16), a prospective population-based cohort study. The participants underwent a non-contrast CT scan between 2003 and 2006, as part of a study on visualization of arterial calcification. Subsequently, the participants were continuously monitored for incident stroke until January 1, 2016. Further details on the follow-up procedures can be found in the Appendix E1 (supplement). The flow of participants and data through our study, with exclusion criteria, is shown in Figure 1.

The Rotterdam Study (16) has been approved by the Medical Ethics Committee (registration number MEC 02.1015) and by the Dutch Ministry of Health, Welfare and Sport (Population Screening Act WBO, license number 1071272-159521-PG). All participants provided written informed consent to participate in the study and to have their information obtained from treating physicians.

### Image Acquisition

Scans used for ICAC assessment were acquired using a 16-slice or 64-slice multidetector CT-scanner (Somatom Sensation 16 or 64, Siemens, Forchheim, Germany). Scan parameters were: 16 mm × 0.75 mm collimation, 120 kVp, 100 effective mAs, 0.5 sec rotation time and normalized pitch of 1. Images were reconstructed with in-plane resolution of 0.23 mm × 0.23 mm, slice spacing 0.5 mm, slice thickness 1 mm and 120 mm field of view.

## Manual ICAC Assessment

ICAC was segmented bilaterally in the intracranial internal carotid artery from its horizontal petrous segment until the circle of Willis by two physicians (DB [3 years of experience] and R2 [1 year of experience]). Both readers were blinded to participant data. The observers were trained and supervised exclusively in evaluating ICAC by an expert neuroradiologist (more than 15 years of experience), whom they could consult in case of doubt while annotating the data. Most scans were segmented by only one observer. A total of 47 randomly selected scans were segmented by both observers independently to assess the interobserver agreement. The segmentation was performed by manually circling calcifications in every section and subsequently selecting pixels with attenuation above 130 HU. This assessment method has been used extensively over the past years with a high intra- and inter-observer reliability (2, 7-8, 17-20).

## Automated ICAC Assessment

The pipeline of the automated method is shown in Figure 2. The scan was first automatically preprocessed (as described below) and then was processed by an ensemble of four deep learning networks. Each network has the same architecture similar to U-Net (21) and, to increase diversity in the ensemble (22-23), was trained individually using a different loss function: cross-entropy (10), Dice overlap (24), focal loss (25), or a weighted cross-entropy loss up-weighing smaller calcifications (descriptions are in Appendix E2 [supplement]). The ensemble produced four probability maps representing networks' confidence in classifying pixels as ICAC. The four maps, corresponding to the four networks, were averaged to obtain the final probabilistic segmentation map. The ICAC volume was computed as the number of pixels with probability above 0.5 multiplied by pixel dimensions and slice spacing.

## Data Preprocessing

We aligned all scans to a single arbitrarily chosen reference image using affine image registration with the default similarity metric ("AdvancedMattesMutualInformation") using SimpleElastix toolbox (26) and cropped them along the longitudinal axis so that they contain only the intracranial part of the carotid artery. These steps were done completely automatically. Then, we reduced the resolution of axial slices so that it matched that of the longitudinal axis. The same transformation was performed on the segmentation maps. These three steps were aimed at reducing the size of the input for deep learning networks, which was necessary to cope with the restrictions imposed by limited computational capacities, particularly GPU memory. The resulting image size was $240 \times 240 \times 100$ and each pixel corresponded to $(0.5 \text{ mm})^3$. We used these registered, cropped and downsized CT volumes and segmentations both to train and evaluate our networks.

## Deep Learning

The choice to use multiple networks instead of one was motivated by the fact that ensemble methods have been shown to boost performance and increase stability of predictions (22), especially when the ensemble is diverse (23). To increase the diversity, the ensemble networks were trained using different objective functions, each weighing the importance of different pixel categories differently. Table E1 (supplement) reports the segmentation performance of individual networks in our ensemble (named after objective functions they use) and compares it to the performance of the entire ensemble. The ensemble achieved higher overall performance compared to its individual networks.

The network architecture, shared by all ensemble members, was similar to our previous version of the method (15) (depicted in Figure 1 of the cited manuscript). However, we made several simplifications: we removed auxiliary classifiers, dropout layers, residual connections, and convolutional layers in concatenation blocks. The member of our ensemble trained using cross-entropy loss function is thus the most similar to our previous version of the method (15), which was a single network that used the same loss.

The rest of the training parameters were as follows. Batches consisted of two large patches of size 178 × 178 × 98 sampled from the scans. The training duration was 75 epochs, with an epoch defined as iterating through 750 training and 1000 validation patches. The lowest validation loss was used to select network parameters. Adadelta algorithm was used for optimization.

Ten-fold cross validation was used for model evaluation. Approximately 1,590 CT scans were used for training and 500 for validation in every cross-validation fold; the same data split was used for all four models. Each fold had approximately 230 test CT scans.

The method was implemented in Python using deep learning framework Keras with Tensorflow backend.

## Visual Assessment of Segmentations

We randomly sampled 300 two-dimensional image regions centered at either the left or right intracranial carotid artery and having the difference between manual and automated segmentations of at least 2 mm$^2$. An expert reader (neuroradiology and head and neck radiology, 8 years of experience), indicated the following for every region: *(a)* whether the manual or automatic segmentation contour (if any) was more accurate and to what extent (slightly, substantially, or equal; five categories in total); *(b)* whether the visualization permitted assessment of ICAC; and *(c)* whether at least one of the contours was accurate. The visualization technique for presenting segmentations blinded the observer to whether the contours had been generated manually or automatically. See the Appendix E3 for details regarding the sampling and visualization and Supplemental Videos E1 and E2 for demonstrations of the visualization.

## Statistical Analysis

We evaluated the accuracy of automatic segmentations using the observer segmentations as the reference standard. The performance metrics used were: recall (sensitivity), precision (positive predictive value), and false positive volume (FPV: volume corresponding to non-ICAC method-detected pixels). In addition, we computed precision-recall and free-response receiver operating characteristic curves.

To assess the variability between manual and automatic volume measures, we used Spearman and intraclass correlations (ICC(2, 1) in the Shrout and Fleiss convention (27)), as well as Bland-Altman analysis (28).

We used Cox proportional hazards models to relate manually and automatically assessed ICAC presence and volume to incident stroke, adjusting for age, sex, scanner type, obesity, hypertension, diabetes mellitus, hypercholesterolemia, low high-density lipoprotein cholesterol, and smoking.

Automatic segmentations and volume measurements used to compute performance metrics and association measures were computed in a ten-fold cross-validation procedure (see Figure 1). All metrics and association measures we report were computed using automated assessments on independent test sets (ie held-out, unseen during training) from the cross-validation procedure; scans from all ten test sets were combined into one set (unless otherwise specified) to compute the metrics. All statistics were computed using Python package SciPy 1.0, IBM SPSS Statistics 24, and R 3.2.3. Significance threshold was 0.05.

# Results

## Participant Overview

A total of 2,319 participants were included for the development of the ICAC segmentation and volume measurement model. The mean age at the scan time was $70 \pm 7$ years, 1154 (53.2%) participants were women, and 1486 (69%) were scanned using the 64-slice scanner. In the set of 47 participants for which ICAC was annotated by two observers independently, the mean age was $67 \pm 5$,

21 (45%) participants were women, and 41 (89%) were scanned using the 64-slice scanner. Table 1 shows clinical characteristics of a subset of participants included into stroke association analysis (White, no prevalent stroke at the scan time).

## Model Training

The time required to train one network (out of four) in the ensemble was approximately one day. The average time the method took to process one scan was 118 seconds (98 seconds for pre-processing, 20 seconds for applying the ensemble and combining its predictions) using Intel Xeon E5645 processor (6 cores, 2.40 GHz) and Nvidia GeForce GTX 1070 graphics card.

## Segmentation Performance

Figure 3 shows a comparison of the performance of the automated method with that of the observers using precision-recall curve and receiver operating characteristic curve analysis. All performance metrics for assessing segmentation performance of the method and the observers are summarized in Table 2.

To assess the training stability of the method, we computed standard deviations of all metrics across the cross-validation folds. The standard deviation for dataset-wise recall was 83.8 ± 1.8%, precision 88 ± 1%, participant-wise recall 80.6 ± 1.7%, FPV among participants with ICAC 15.7 ± 2.4 mm$^3$, and FPV among ICAC-free participants 6.2 ± 2.4 mm$^3$, indicating that the method provided consistent performance when trained on different data subsets.

## Volume Measurement Performance

The intraclass correlations between the automatic and manual volume measures and between the two observers' measures were 0.98 (95% CI: 0.97; 0.98) and 0.91 (95% CI: 0.84; 0.95; *P* = .04). The corresponding Spearman's correlations were 0.95 (95% CI: 0.84, 0.96), and 0.97 (95% CI: 0.92, 0.99;

*P* = .38). The CIs for Spearman's correlations and significance tests for both correlations were computed by bootstrapping with 10,000 replications.

Figure 4 shows Bland-Altman plots of the difference between automatic and manual volume measures and the difference between cubic roots thereof.

## Visual Assessment of Segmentations

Six out of 300 regions could not be graded either because the observer could not infer the region orientation from the visualization due to its limited scope, or because of the limited spatial resolution. Table 3 presents visual comparison of accuracy of automatic and manual segmentations for the following subsets of regions*: (a)* all gradable regions*, (b)* regions in which the method did not miss any ICAC pixels ('false positives'), and *(c)* regions in which it did not segment any pixels not indicated in the manual segmentations ('false negatives'). Manual and automatic segmentations were both inaccurate in only six of 294 gradable regions.

## ICAC and Incident Clinical Stroke

We assessed associations of presence and volume of ICAC with stroke incidence. For both ICAC presence and volume measures, stroke associations of manual and automated assessments were similar. Adjusted hazard ratios computed for ICAC presence were 2.51 (95% CI: 1.42, 5.85) for automated and 2.52 (95% CI: 1.44, 5.95) for manual presence assessment; P = .99. Adjusted hazard ratios per one-standard-deviation increase of measured ICAC volume were 1.38 (95% CI: 1.12, 1.75) for automated and 1.48 (95% CI: 1.20, 1.87) for manual volumes; P = .12. The CIs and significance tests were computed by bootstrapping with 10,000 replications. We explore differences in volume and attenuation distributions between automatically and manually identified ICAC lesions in Appendix E5 and investigate whether these differences may affect the association of automatically computed ICAC volumes with stroke in Appendix E6.

# Discussion

We developed a fully-automated deep-learning-based method for ICAC segmentation and volume measurement in non-contrast CT and evaluated it on a large dataset in a ten-fold cross-validation procedure. Accuracy of automated assessment was comparable to or better than manual assessment by trained observers in several aspects: segmentation and volume measurement performance measured between automated and manual assessments were comparable to those measured between the observers; in blinded visual assessment of segmentations, automated segmentations were more accurate than manual (131 of 225 regions with automated and manual delineations judged as not equally accurate; 58.2%; $P = .01$).

Furthermore, the method identified lesions missed by the observers, while detecting few non-ICAC structures. In 77% (78 of 101) of 'false positive' regions the automated segmentation was more accurate than manual, the opposite happened in only 16% (16 of 101) of cases (see Table 3). However, the pattern was different in 'false negative' regions: when the method missed observer-indicated ICAC it was wrong in 49% (67 of 138) and correct only in 28% (38 of 138) of the cases. This suggests that a union of automated and manual segmentations may yield a more accurate measurement of ICAC volume.

The associations with stroke for the automated presence and volume assessments were similar to those for the corresponding manual assessments. However, our analyses in Appendices E5 and E6 suggest that there may be differences in kinds of lesions the method and the observers focused on. More specifically, the method detected more small and low-attenuation lesions than the observers and that could possibly affect the stroke association for automated volume measurements. Nevertheless, ICAC volumes were found to be associated with stroke, and thus could be useful for stroke risk assessment models. For example, automated assessment might be used in primary stroke prevention in the future, similarly to how coronary artery calcium assessment is currently used to re-classify persons at risk of a first cardiac event (29). We would like to emphasize that the purpose of analyzing stroke associations was to further study potential differences between manual and automated measurements and to assess the potential of the latter for use in stroke risk estimation; the purpose was not to demonstrate the association between ICAC and stroke, which was studied before (2).

Apart from automated assessment, another cheaper alternative to manual segmentation-based ICAC assessment is visual scoring, in which an observer assesses ICAC severity subjectively. Scoring systems categorizing entire scans or arteries into severity grades have substantial (7) to excellent (18) interobserver agreements, but show very large variations of ICAC volume within grades. Ahn et al (18), who compared several such systems, concluded that the segmentation-based volume measure is more promising for assessing arteriosclerosis. A section-wise scoring method of Subedi et al (6), shown to be superior to scan-wise scoring, achieved Spearman's correlation of 0.91, whereas for our method this was 0.95. Automated assessment accurately emulates segmentation-based volume measurements and, unlike visual scoring, does not require any human input, which may make it a better candidate for replacing manual assessment.

The main limitation of the current study is that the method was trained and evaluated on a relatively homogenous dataset: scans of (mostly White) persons from one country acquired using a standardized protocol on two scanners of one vendor. Evaluating the method on scans of trauma patients reconstructed using a different kernel (see Appendix E4 and Figures E2 and 5) showed that, although the agreement with manual annotations was good (intraclass correlation was 0.94 and Spearman's correlation 0.95), it was lower than that on the Rotterdam Study data, used for training. To generalize our model to data which may be dissimilar from our training data, such as the trauma dataset, the training data could be expanded using scans from the target population of scans (ie the population we wish to apply the method to) or other type of scans dissimilar to the training data (which would make the training data more heterogenous and thus may improve generalization). Alternatively, the method itself could be augmented to adapt to new data, or the target data could be transformed to be more similar to the training data.

Another possible limitation is that the quality of manual segmentations used in training and evaluation could have been reduced due to labor-intensiveness and tediousness of the annotation process and/or that the annotators did not specialize in neuroradiology. Having a larger number of scans annotated by several observers, including expert neuroradiologists, could allow training a better algorithm and a more comprehensive, accurate comparison between automated and manual assessment.

In this study, we have demonstrated an accurate and fast automated method for ICAC assessment. Automated assessment may replace manual assessment in research or clinical settings, facilitating analysis of large amounts of data. Automated assessment could also be used as a starting point for manual annotations, which would speed up annotation and could increase completeness of ICAC burden estimation. The latter use case may also increase completeness of ICAC burden estimation. Automated assessment may thus facilitate studying the etiology and clinical consequences of arteriosclerosis, for which ICAC is a proxy. It may thus play an important role in providing a basis for and facilitating the incorporation of ICAC-based imaging markers into clinical practice: for example, ICAC volume may be used in stroke risk assessment.

# Figures

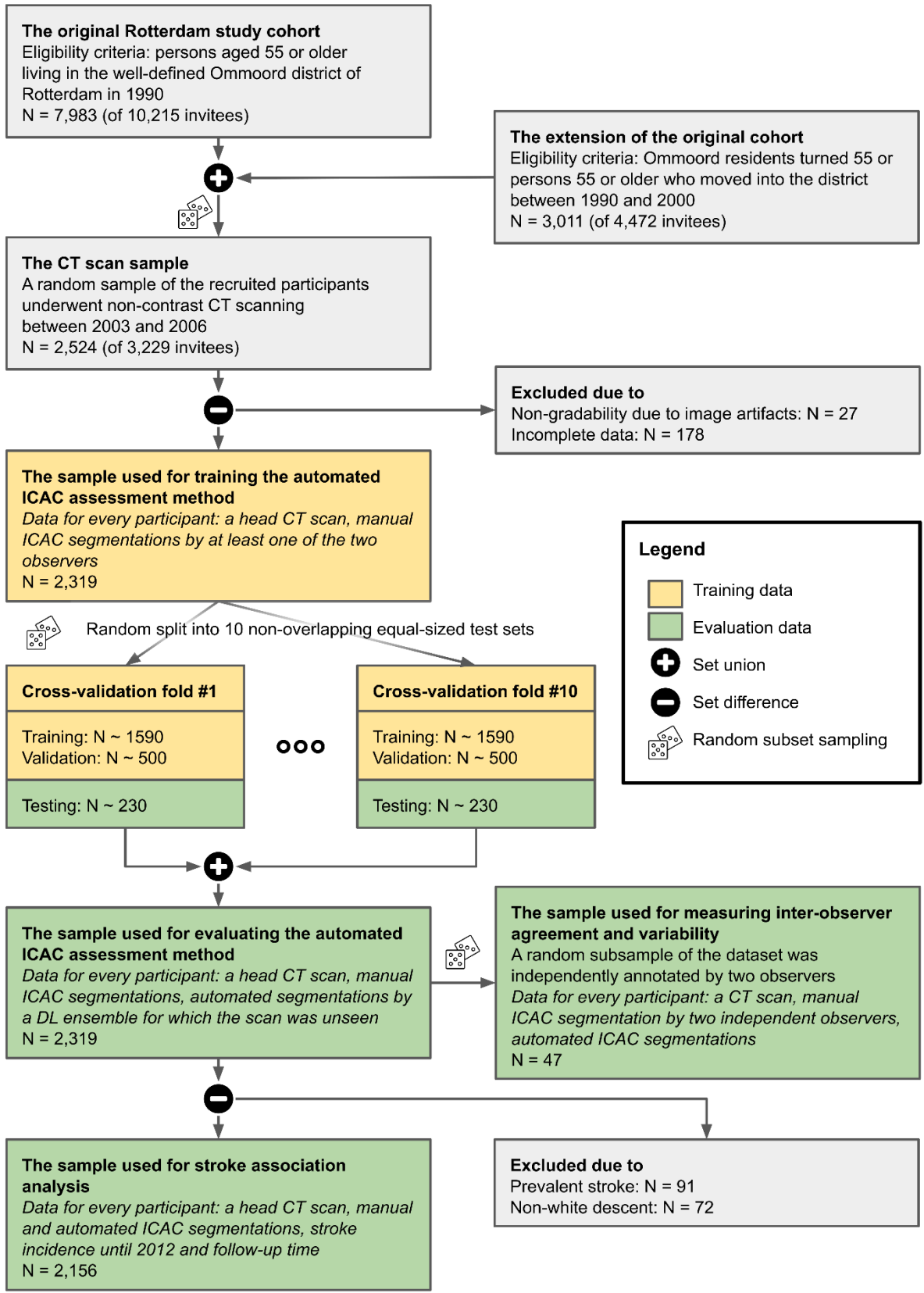

**Figure 1:** The flow of participants and data through the study. In this study, we first selected participants for which both head CT scans and manual ICAC assessments were available (*n* = 2,319; only one scan was available per participant). Participants were randomly partitioned into ten approximately-equal-sized non-overlapping subsets. Ten-fold cross validation was performed by training the model on each of the ten training and validation sets obtained by exclusion of corresponding test dataset. Finally, the test sets from cross-validation were aggregated into a dataset of 2,319 scans with both manual and automated assessments, which were used to evaluate automated assessments against manual ones. In addition to this set, we also performed analyses on its two subsets: a subset for which independent assessments of two observers were available and a subset we used for stroke association analysis and formed using inclusion criteria from a previous study (2).

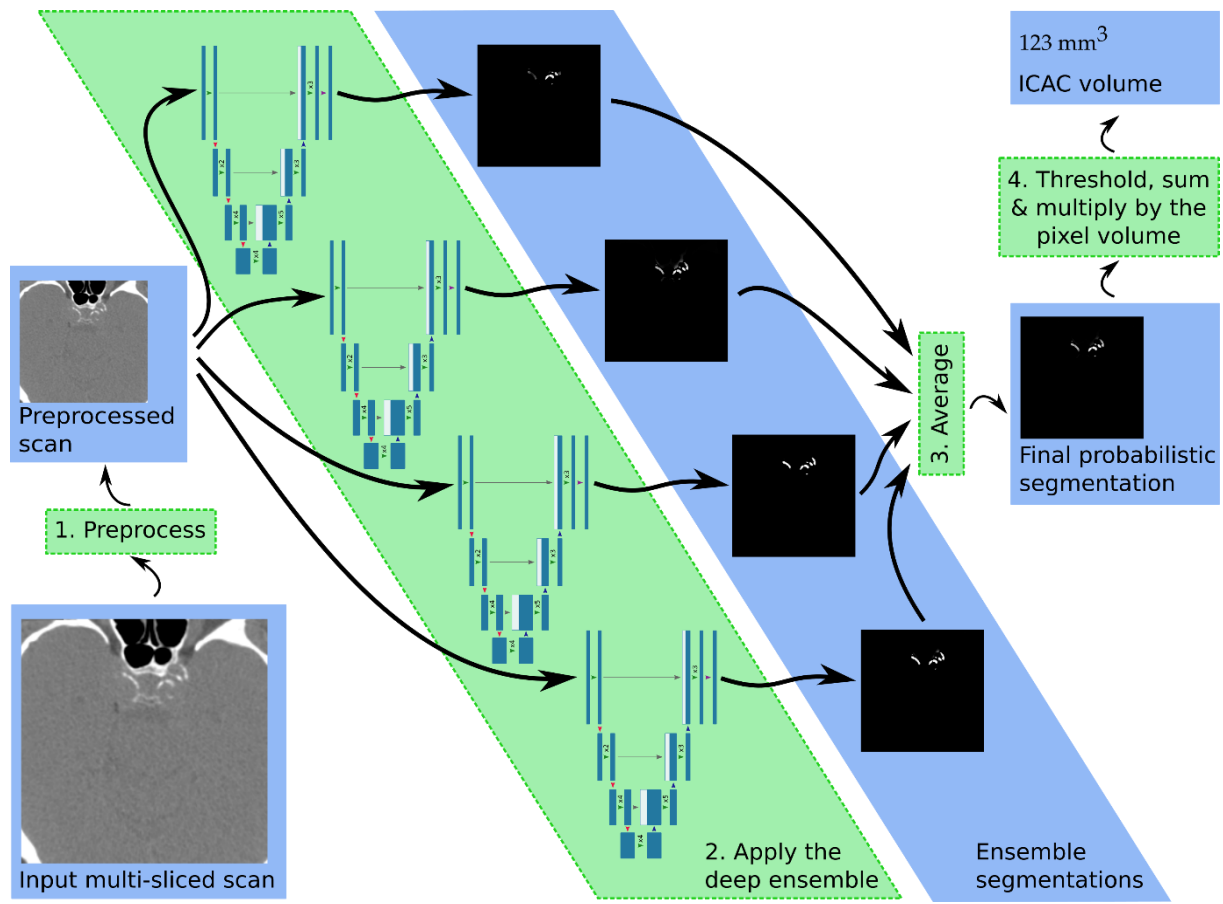

**Figure 2:** The automated method's processing pipeline. The steps: 1) preprocessing; 2) processing by four trained networks ('deep ensemble'), outputting ICAC probability maps; 3) averaging the maps, 4) computing the volume corresponding to pixels with the probability above 0.5.

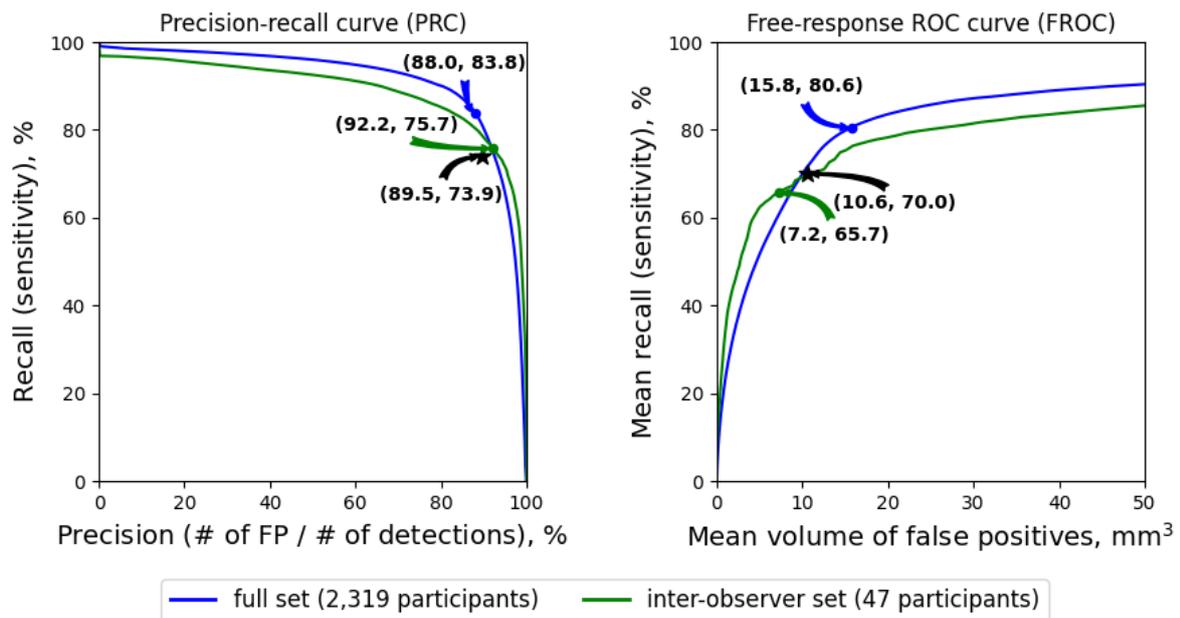

**Figure 3:** Automatic and manual intracranial carotid artery calcium (ICAC) segmentation performance. The performance was evaluated on all scans (blue curve) and those annotated by both observers (green curve). The curves were computed by varying the threshold for ICAC probability maps (from 0 to 1); every point thus represents recall and precision (for precision-recall curve [PRC]) or FPV (for free receiver operating characteristic curves [FROC]) computed over all pixels in all scans (for PRC) or averaged among participants with ICAC (for FROC) using a specific threshold. Stars represent the inter-observer agreement: dataset-wise (ie across all pixels in all scans, averaged-per-participant recall and precision with FPV of R2, with DB used as the reference standard). Dots represent the model performance when 0.5 was used as the probability cut-off. The inter-observer agreement points lie on or under both PRC and FROC, which indicates that the method segmentations agreed with DB at least as much as R2, when an appropriate threshold was chosen. ROC = receiver operating characteristic, FP = false positives.

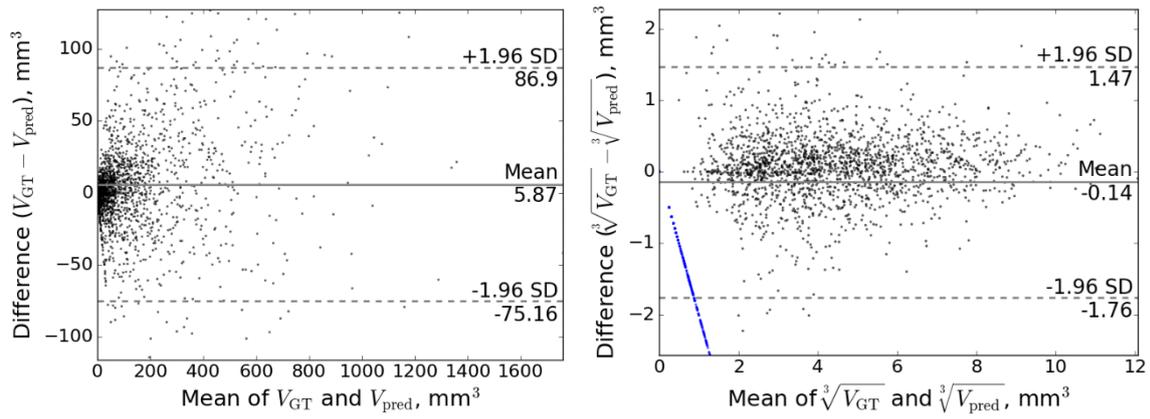

**Figure 4:** The Bland-Altman plots of the difference between manual and automatic intracranial carotid artery calcium volume measures: **(a)** The difference between manual and automatic volumes ($V_{GT}$ and $V_{pred}$) and **(b)** the difference between the cubic root thereof. Blue dots represent participants with $V_{GT}$ of zero (constituting 18.2% of the dataset).

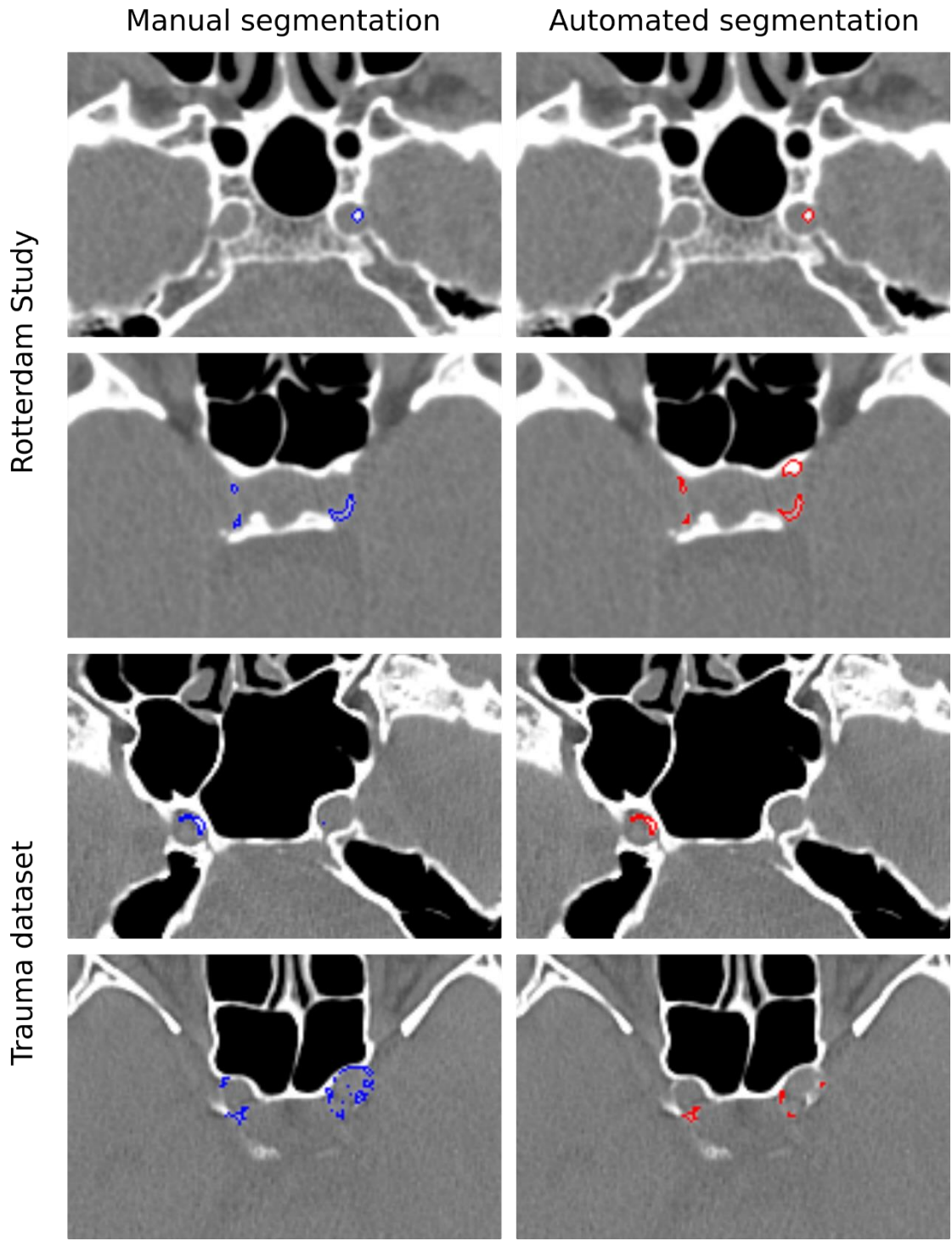

**Figure 5:** Examples of manual and automated segmentations of Rotterdam Study and trauma dataset scans. The images shown were preprocessed (registered and downsized; details on preprocessing can be found in the corresponding subsection of the Materials and Methods). Trauma dataset images were additionally smoothed with a Gaussian filter to make them more similar to Rotterdam Study scans used for training and thus improve the method's performance. Analysis of the method's performance on the trauma dataset can be found in Appendix E4.

# Tables

**Table 1: Baseline Characteristics of the Study Population**

| Characteristic | Value |
|---|---|
| **Sample size** | 2156 |
| **Women** | 1154 (53.5%) |
| **Scanner type** | |
|    16-section | 670 (31.1%) |
|    64-section | 1486 (68.9%) |
| **Age, years** | 69.3 ± 7 |
| **Obesity** | 510 (23.7%) |
| **Hypertension** | 1586 (73.6%) |
| **Diabetes** | 257 (11.9%) |
| **Hypercholesterolemia** | 1064 (49.4%) |
| **Low high-density lipoprotein** | 264 (12.2%) |
| **Past or current smokers** | 1483 (68.8%) |
| **Presence of ICAC** | 1748 (81.1%) |
| **ICAC volume*, mm$^3$** | 42.9 (6.3-139.7) |

Note.—Values are mean (SD) for continuous variables, participant number (percentage) for dichotomous variables.

* Median (interquartile range).

**Table 2: Automatic and Manual ICAC Segmentation Performance**

| Model or reader | Model (*n* = 2,319) | Model (*n* = 47) | R2 (*n* = 47) |
|---|---|---|---|
| **Reference segmentation** | DB + R2 | DB | DB |
| **A. Dataset-wise statistics** | | | |
| **Recall** | 83.8% | 75.7% | 73.9% |
| **Precision** | 88.0% | 92.2% | 89.5% |
| **B. Participant-wise statistics** | | | |
| **With ICAC (*n* = 1897)*** | | | |
|   Recall | 80.6% ± 20.8% | 65.7% ± 29.7% | 70.0% ± 29.8% |
|   FPV, mm³ | 15.8 ± 24.5 | 7.2 ± 17.3 | 10.6 ± 20.0 |
|   Volume, mm³ | 150.3 ± 204.6 | 121.9 ± 191.7 | 121.9 ± 191.7 |
| **ICAC free (*n* = 422)*** | | | |
|   FPV, mm³ | 6.2 ± 12.8 | 2.1 ± 2.6 | 0.0 ± 0.0 |

Note.—Participant-wise values are mean ± standard deviation. DB and R2 are observers. The metrics the second and third columns were computed using the exact same set of reference segmentations. 'Volume' is the intracranial carotid artery calcification (ICAC) volume for participants with ICAC as assessed by the observers, reported for comparison with FPV. FPV = false positive volume.

* As indicated by the observer(s).

**Table 3: Blinded Qualitative Visual Comparison of Manual and Automatic Segmentations by an Expert**

|  | All $n = 294$ | False positives $n = 101$ | False negatives $n = 138$ |
|---|---|---|---|
| **Automatic is substantially more accurate** | 14 (4.8%) | 9 (8.9%) | 5 (3.6%) |
| **Automatic is slightly more accurate** | 117 (39.8%) | 69 (68.3%) | 33 (23.9%) |
| **Both are equally accurate** | 69 (23.5%) | 7 (6.9%) | 33 (23.9%) |
| **Manual is slightly more accurate** | 86 (29.2%) | 16 (15.9%) | 59 (42.8%) |
| **Manual is substantially more accurate** | 8 (2.7%) | 0 (0%) | 8 (5.8%) |
| ***P* value** | < .001 | < .001 | .008 |

Note.—Values are region number (percentage). P-values were computed using Wilcoxon signed-rank test and represent the significance of the difference between accuracy of manual and automatic assessments. N is the number of regions in the given category.

# Supplemental Material

## Appendix E1. Follow-up Procedures for Stroke

All participants who underwent the CT-scan were continuously monitored for incident stroke through linkage of the study database with files from general practitioners. Additional information was obtained from hospital records. All potential strokes were reviewed by research physicians and verified by an experienced stroke neurologist.

## Appendix E2. Loss Functions

The four different objective functions (also known as "loss functions") used to train the ensemble are described; similarly to our previous study (15), we only defined these objective functions on pixels with attenuation above 130 HU. First, standard cross-entropy (10), weighs all pixels the same. Second, scan-wise soft (differentiable) Dice loss (24) optimizes for a good balance between true and false positive rates on the scan level and consequently emphasizes correct classification of ICAC pixels in scans with low ICAC volume (ie low number of ICAC pixels). Third, recently developed focal loss (25), up-weighs pixels with low classification confidence, since they are likely to be 'hard' examples. Fourth, a custom weighted cross-entropy loss that up-weighs pixels depending on the size of lesion they belong to: pixels in lesions of size 10 to 100 pixels have weight linearly decreasing from 10 to 1, the rest of the pixels have a weight of one.

## Appendix E3. Visual Assessment of Segmentations

The regions for visual assessment were sampled as follows. First, regions were obtained from all scans by cutting axial slices in half along the sagittal plane, so that each region contained either the left or right intracranial carotid artery. Second, 300 regions were sampled uniformly at random from regions in which the difference between manual and automatic segmentations (false negative plus false positive area) was at least 2 mm². Only one region was sampled per participant to ensure diversity.

Figure E1 shows examples of analyzed regions as they were presented to the observer. The regions were visualized by sequences of images showing manual and automatic segmentation contours overlaid on the target scan slice and its 10-slice neighborhood. The visualized region area was re-centered around calcifications and corresponded to a $50 \times 50$ mm² area. The slices were visualized using window level and width of 40 and 850, respectively. The contour positions (left / right) and colors (blue / red) were random so as to blind the observer to how the contour was obtained (manually or automatically). Visualizations of 20 randomly sampled analyzed regions, using random swapping (exactly as they were presented to the observer) and not using it, and the corresponding expert's assessments thereof are provided as Supplemental Videos E1 and E2, respectively.

# Appendix E4. Evaluation of Automated ICAC Assessment on An Independent Dataset

To further assess the generalization capability of the proposed method, we evaluated it on an external dataset of 125 non-contrast thin-slice CT scans of the skull acquired from trauma patients.

## Setting

In this evaluation, we used a random sample of a dataset originally collected to study prevalence and distribution of ICAC across different age groups (30). The data was acquired from a random sample (same number of patients was sampled per age decile per sex) of trauma patients who were admitted to the emergency department of Erasmus MC between 2009 and 2016 and underwent non-contrast thin-slice CT of the skull. After excluding patients with severe intracranial hemorrhage, suspected stroke, and those whose scans showed severe artefacts, the resulting study sample contained 868 patients. The study was approved by the institutional review board. Further details on this study can be found in (30). In the present analysis, we randomly sampled 125 participants of age 55 and above

(to match the inclusion criterion of the Rotterdam Study) for which ICAC was annotated in scans and for which reconstructions with the same kernel were available. Only one scan per patient was used. In this sample of patients, the mean age was 74.4 (SD 8.9) 48.8% patients were women.

## Image Acquisition

The scans were acquired using Siemens Somatom Definition Edge scanner (Forchheim, Germany). The scanning and reconstruction parameters varied. In the sample used for evaluating the proposed method, in-plane resolution ranged between 0.39 mm × 0.39 mm and 0.49 mm × 0.49 mm, slice spacing between 0.4 and 1.0 mm, slice thickness between 0.75 mm and 1 mm. All scans were reconstructed using the sharp 'J70h' convolution kernel. Examples of images from the Rotterdam Study and trauma dataset are given in Figure E2.

## Manual ICAC Assessment

ICAC was annotated in every slice by two medical students with two years of experience, trained by an expert. The annotations were performed using the same semi-automated technique as described in the Manual ICAC Assessment subsection of the manuscript.

## Automated ICAC Assessment

Pre-processing of the trauma dataset was different from preprocessing of the Rotterdam Study data (see Supplemental Materials and Methods for a detailed description) in a few ways due to the fact that trauma dataset scans have different field of view and capture a somewhat different anatomic region (e.g. trauma dataset scans usually capture the entire skull, see Figure E2). Firstly, before the registration step, the images were resized, re-centered in the axial plane (around the center of mass of voxels above 0 HU), and cropped to match the field of view of the Rotterdam Study scans. Secondly, the number of optimization steps for image registration was enlarged from 128 to 1024.

The remaining differences between the evaluation on the Rotterdam Study and trauma dataset are as follows. Registration has failed in a 11 out of 125 (8.8%) cases and these scans were excluded from further assessment, leaving 114 scans. Failed registration was detected automatically by using as a criterion a mean pixelwise absolute error of above 300 HU between the reference image and the image to be registered. Due to the reconstruction kernel used to reconstruct trauma dataset scans ('J70h') yielding substantially more noisy images than the one used in Rotterdam Study scans ('B35f'), which can be seen in Figure E2, the images were smoothed before applying the deep ensemble. Examples of the method's segmentations without smoothing are shown in Figure E3. The degree of smoothing was tuned on a random sample of 14 images. We tried 2D Gaussian kernels with sigma varying between 0.1 and 1.0 with an interval of 0.1. Sigma of 0.6 yielded the best results. Examples of the method's segmentations with the best degree of smoothing are shown in Figure E4. Removing pixels with attenuation below 130 HU in the original versions of images and pixels with attenuation below 130 HU in the smoothed versions of images from final automated segmentations yielded better results than using either only original or only smoothed images to remove below-130HU pixels. From manual segmentations, only pixels below 130 HU in original images were removed.

The performance was evaluated on the remaining 100 scans not used for tuning. The segmentation performance metrics for both datasets are reported in Table E2. For the trauma dataset, the results on both smoothed and original images are provided (example segmentations are shown in Figures E3 and E4).

The volume measurement performance metrics were as follows. For the trauma dataset, when smoothing was used, intraclass correlation was 0.94 and Spearman's correlation 0.95. When smoothing was not used, intraclass correlation was -0.11 and Spearman's correlation 0.58.

## Discussion and Conclusion

The performance of the method on the trauma dataset was rather good, although less good than the performance on the Rotterdam Study dataset, on which the method was trained. We can suggest several ways to improve the performance on data substantially dissimilar to our current training data (including data similar to the trauma dataset). Firstly, if the method is planned to be applied to a specific, known in advance data distribution, and labels for a subset of this data is available, this data can be added to the training and validation sets for the method and given a sufficient weight during optimization; in this way, the method can adjust to the target distribution. Secondly, if labeled target data is unavailable for training, and/or one would like to build an algorithm that performs well on a larger variety of data, the training data can be extended to cover a larger variety of scanning and reconstruction settings, acquisition sites, participant demographic and clinical characteristics. Thirdly, techniques such as data augmentation (for example, extending the training data by simulating a large variety of contrast changes or added noise) or domain adaptation (e.g., adversarial training (31)) could be used to train the method to be less sensitive to changes without the need for additional data for training. Finally, the data from the new domain could be transformed to be more similar to that from the training domain. We used such a technique to improve the performance of the method on trauma dataset, reconstructed using a sharper kernel: we smoothed the trauma dataset scans with a Gaussian kernel prior to applying our deep ensemble. Other techniques could deal with such a difference between training and testing domains better, for example, machine learning methods for converting CT reconstruction kernels, such as (32), since smoothing cannot accurately simulate this conversion.

## Appendix E5. Volume and Attenuation Distributions of ICAC Lesions in Manual and Automated Segmentations

Figure E5 presents 2D histograms of volume and attenuation of lesions identified by manual and by automated annotation, and that for false positive lesions (lesions detected by the automated method but not annotated by the observers) and false negative lesions (lesions identified by the observers but missed by the method). The details of the computation of these histograms are as follows. The bin ranges for volume and attenuation were defined as percentiles of volumes and densities of all manually identified lesions, respectively, adjusted for volume (for example, lesions below 145 HU — $1^{st}$ percentile of attenuation — constitute 1% of total ICAC volume in all participants). Individual lesions were identified and quantified as follows. The lesions were identified by connected-component labeling of manual and automated segmentations in 3D. The volume of a lesion was computed as a volume of the respective connected component and the attenuation was a median of densities of pixels constituting the lesion.

As can be seen from Figure E5-C, most false positive lesions had low volume and attenuation: lesions with volume above 10 mm$^3$ ($10^{th}$ percentile of all lesions) and attenuation above 183 HU ($10^{th}$ percentile of all lesions) constituted only 24% of total volume of all false positive lesions.

The distribution of false negatives appeared to be more spread-out than that of false positives (Figure E5-D): lesions with volume and attenuation above the corresponding 10th percentiles constituted 44% of the volume of all false positive lesions. The majority of those had attenuation above the median attenuation — 266 HU.

In summary, there were differences in distribution of manually and automatically identified lesions in terms of their volume and attenuation. Most notably, the automated method identified more small and low-attenuation lesions and less higher-attenuation lesions than the observers.

# Appendix E6. Difference in Stroke Associations for Manual and Automatic ICAC Volumes

In this subsection, we investigate whether potential differences in incident stroke associations for manually and automatically computed ICAC volumes in stroke associations could be (partially) attributed to the differences in distributions of volume and attenuation of lesions constituting manual and automatic volume measurements described in the previous subsection: namely, the proposed method detecting more small and low-attenuation lesions and under-detecting higher-attenuation lesions compared to manual annotation. For that, firstly, we assessed how excluding smaller and lower-attenuation lesions from automated segmentations impacted the association between stroke and ICAC volume measurements computed from these automated segmentations. Secondly, we assessed how adding higher-attenuation lesions that the observers annotated but the method missed ("false negatives") to automated segmentations impacted the association between stroke and volumes computed from these segmentations.

For this analysis, we estimated associations of automated and manual volume measurements with stroke incidence until 1/1/2012. (The figures reported in the manuscript correspond to the full duration of follow-up, until 1/1/2016.) The adjusted hazard ratio (HR) per 1-SD increase of automatically computed ICAC volume was 1.32, 95% CI [1.01; 1.72]. That for the manual volume was 1.48, 95% CI [1.12; 1.95]. The p-value for the difference between these HRs was 0.03. The HRs were computed using Cox proportional hazards model, which included nine other variables apart from ICAC volume (see Methods section, Statistical analysis subsection). The confidence intervals and p-value were computed using bootstrapping with 10,000 replications. The experiments below used exactly the same methodology, apart from that we had to reduce the amount of replications for bootstrapping to 1,000 due to an increased amount of statistical tests that had to be performed.

## The impact of over-detecting small and lower-attenuation lesions

Figure E6 presents hazard ratios (HRs) for manual and automatic volumes from which volume contributions of lesions with volume and attenuation below varying lesion volume and attenuation percentiles were excluded; figure sections C and D show the difference between these modified automated volumes and the correspondingly modified manual volumes and the original manual volumes, respectively, with the significance levels indicated by stars. Excluding lesions with volume below 5 mm$^3$ (4$^{th}$ percentile) or lesions with attenuation below 174 HU (7$^{th}$ percentile) from manual segmentations (or excluding lesions using thresholds below these values) did not yield a significant decrease in HR, suggesting these lesions were not important contributors to positive association between stroke (in the shorter-term follow-up) and ICAC volume. Excluding lesions with volume below a threshold of 3 or 4 mm$^3$ (3$^{rd}$ and 4$^{th}$ percentiles) and attenuation below 162 HU (4$^{th}$ percentile) or lesions with volume below 3 mm$^3$ and attenuation below thresholds between 162 (4$^{th}$ percentile) and 177 HU (8$^{th}$ percentile) from automatic segmentations yielded increased HRs that were not significantly different from those for both manual volumes with lesions of the same volume and attenuation excluded and original manual volumes (p > 0.05). The same held when not all lesions, but only lesions not detected by the observers below these thresholds were excluded.

## The impact of missing higher-attenuation lesions

Figure E7 shows HRs for automatic volumes to which the volume of false negative lesions (from the manual annotations) with volume and attenuation above varying lesion volume and attenuation percentiles were added; figure section B shows the difference between these modified automated volumes and original manual volumes with the significance levels indicated by stars. Adding to automated segmentations false negative lesions with attenuation above the median (50$^{th}$ percentile) — 266 HU — yielded an HR of 1.37, 95% CI [1.06; 1.84], which was not significantly different from that for manual volume (p = 0.06). However, we performed a similar analysis with adding smaller and lower attenuation false negatives to automated segmentations and found that that could also increase HR to a similar amount: for example, adding false negatives with attenuation below 30$^{th}$ percentile yielded HR of 1.37, 95% CI [1.07; 1.86], which was also not significantly different from the manual volume HR (p-value slightly above 0.05).

## Discussion and conclusion

The results in the previous section ("Volume and Attenuation Distributions of ICAC Lesions in Automated and Manual Annotations") show that the automated method detected more lesions that are small and low-attenuation than the observers and detected fewer high-attenuation lesions (see Figure E5, sections C and D). We hypothesized that the association between stroke and automatic ICAC volumes could be lower than that for manual volume (in part) due to these differences in distribution of volume and attenuation of detected lesions. In this section, we attempted to test these hypotheses.

Smaller and lower-attenuation lesions (up to a certain threshold) did not appear to be important contributors to the association between stroke and manually computed ICAC volumes (as assessed for the shorter-term follow-up, restricted until 1.01.2012). At the same time, automatically computed ICAC volumes that excluded such lesions had increased associations with stroke, which were no longer significantly different ($p > 0.05$) from those for manually computed volumes, original or with small and low-attenuation lesions excluded. This suggests that, to an extent, the stroke association for automatic volumes (computed using all detected lesions) was lower than manual in the shorter-term follow-up due to the automated method detecting more smaller and lower-attenuation lesions. However, this analysis is limited by the fact that not all false positives the method detected were ICAC: the increased association observed when excluding smaller and lower-attenuation lesions from automated segmentations could in part be due to removing non-ICAC structures falsely identified by the method from the volume computations. Since the visual assessment of automated segmentations revealed that only 16% of false positives were not ICAC, we believe this effect was likely not the only reason for the stronger association between stroke and automated volumes after excluding small and low-attenuation lesions.

Although the method missing lesions appeared to weaken the association between automated assessment and stroke, it was unclear whether higher-attenuation lesions contributed more to that than lower-attenuation lesions.

In these analyses, we only looked at the difference between automated and manual ICAC assessments in terms of volume and attenuation of detected lesions. There could possibly be other biases in automated assessment that would further explain the difference in stroke association with manual assessment. For example, there could be differences in location or shape of automatically and manually detected lesions, or differences in accuracy of detecting ICAC in different participant subpopulations, and these differences, in turn, could be related to stroke.

# Supplemental Figures

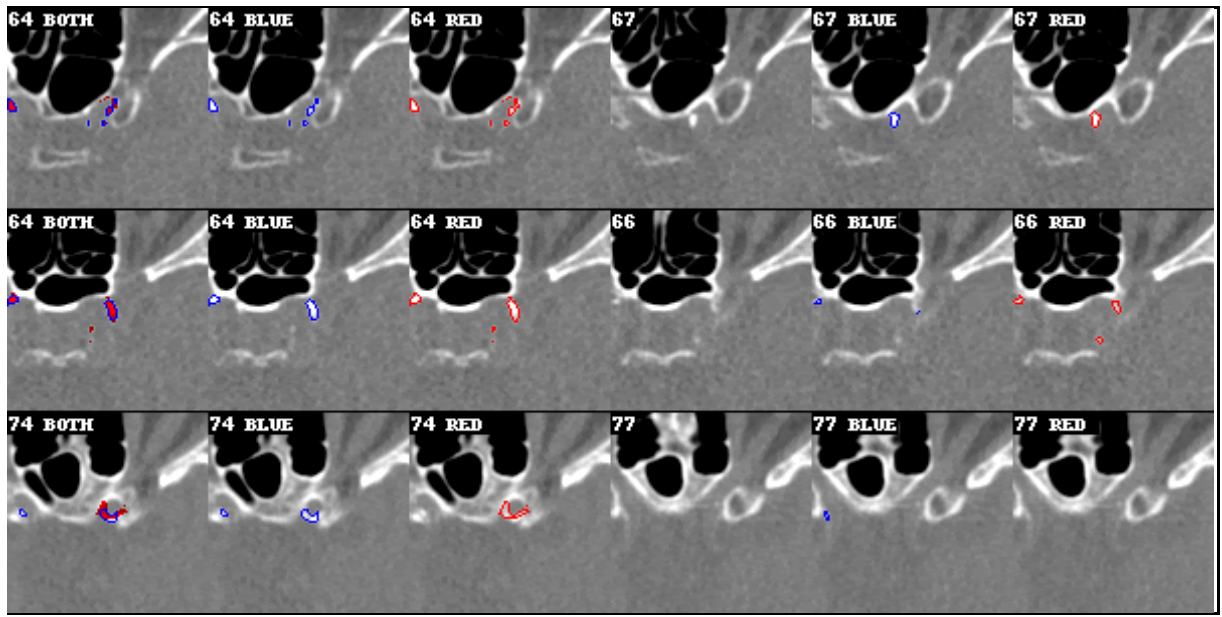

**Figure I: Examples of Regions Visually Assessed by the Expert**

Left to right: images 1 to 3 are the target slice with manual and automatic annotations either superimposed or visualized separately; 4 to 6 show its 10-slice neighbourhood via GIF animation with no annotation or either manual or automatic contours superimposed. The colors (red and blue) and the corresponding screen position (left and right) of the segmentation contours were random in every region so that the observer was blinded to which contour was automatic and which was manual. Top to bottom: blue corresponds to the automatic, manual, and automatic annotations. The regions were labeled by the expert, unaware of the latter information, as 'both are equally accurate', 'red is slightly more accurate' and 'blue is slightly more accurate', respectively.

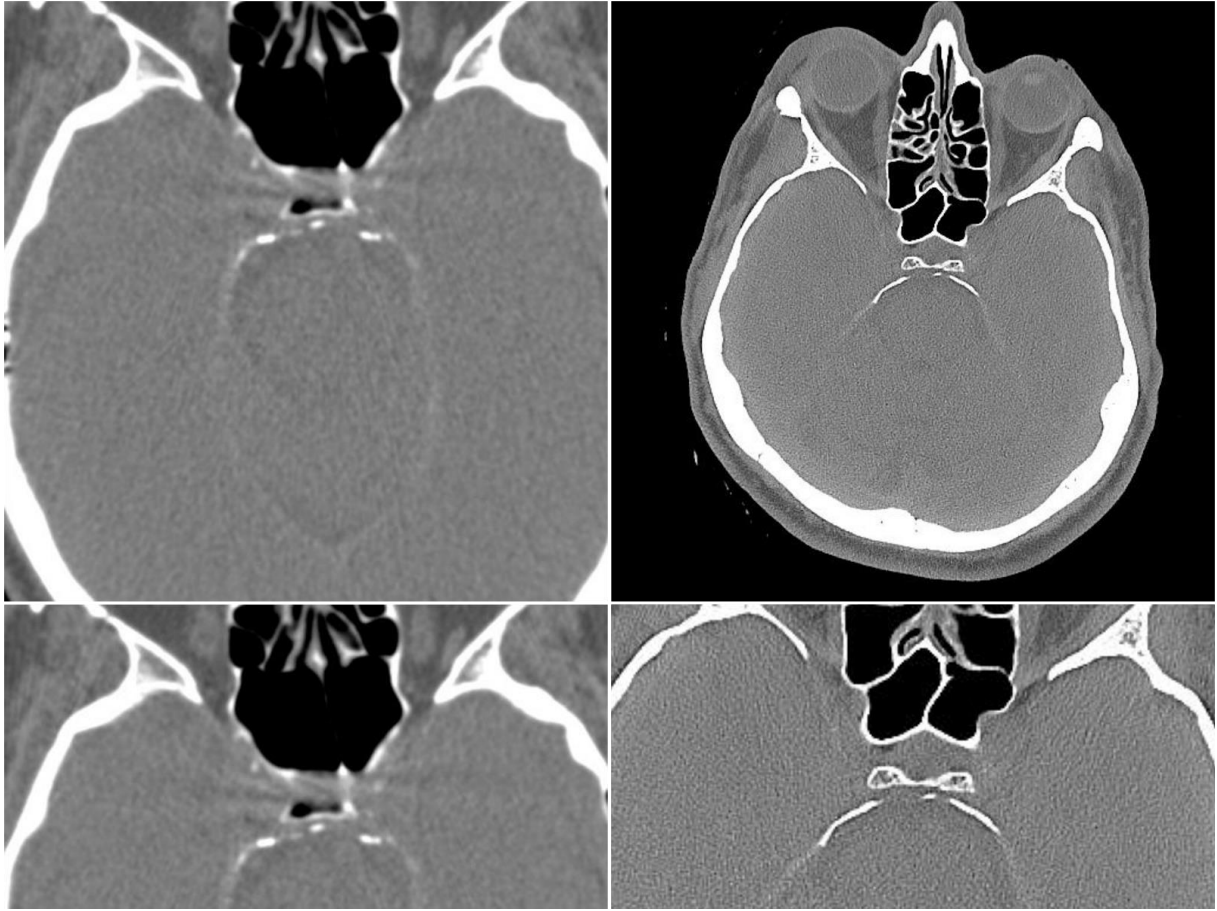

**Figure II: Rotterdam Study and Trauma Dataset Scans**

Example of scans from the Rotterdam Study (left) and trauma dataset (right). Trauma dataset scans have a wider field of view and are reconstructed using a sharper kernel.

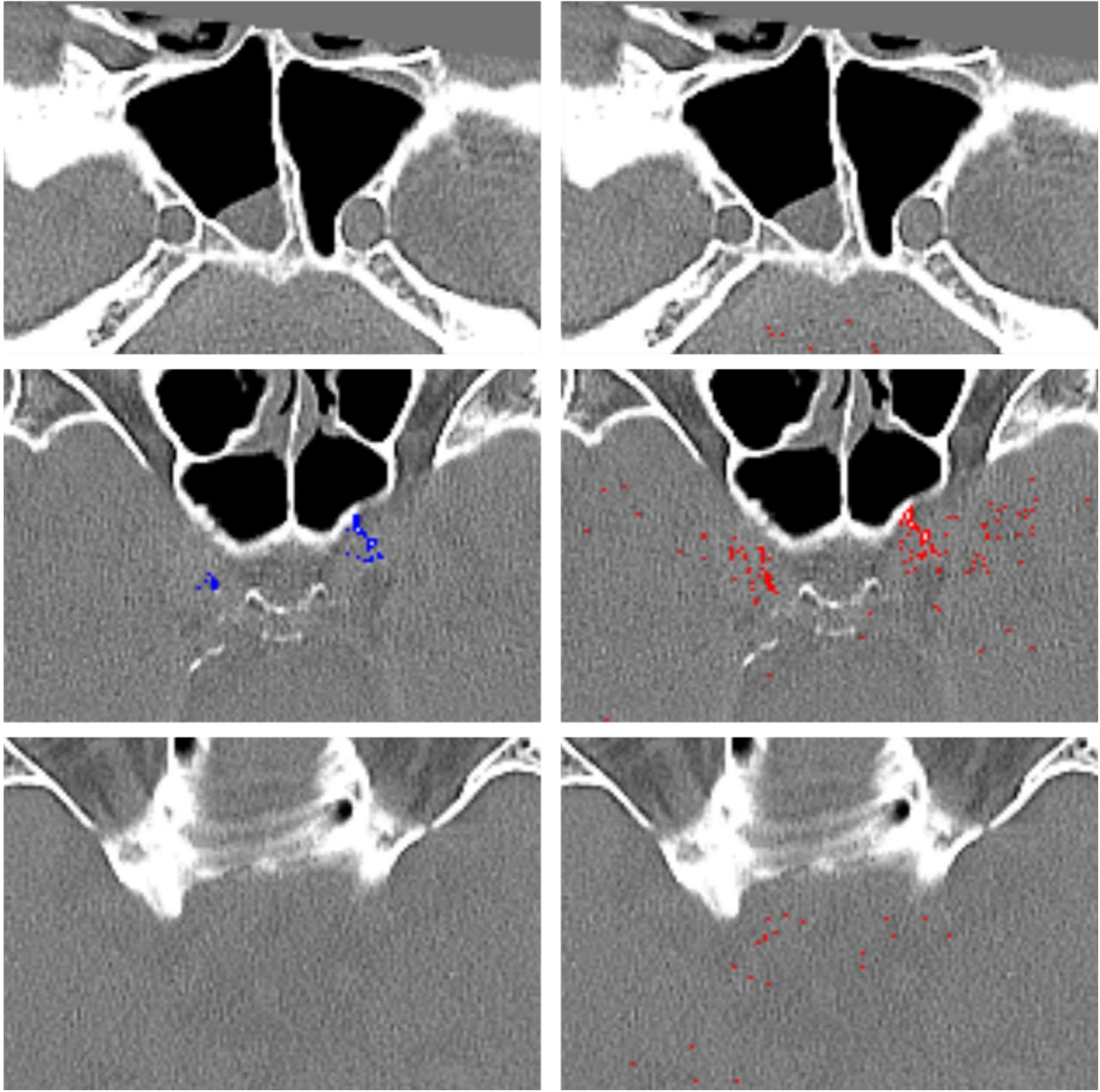

**Figure III: Manual Segmentations and Automated Segmentations Computed Without Pre-Smoothing Input Images**

Manual segmentations (left, blue) and automated segmentations (right, red) computed without smoothing images with Gaussian filter prior to applying the deep ensemble. Without smoothing images prior to applying the deep ensemble, the method produces many false positives, which are often noise with density above 130 HU.

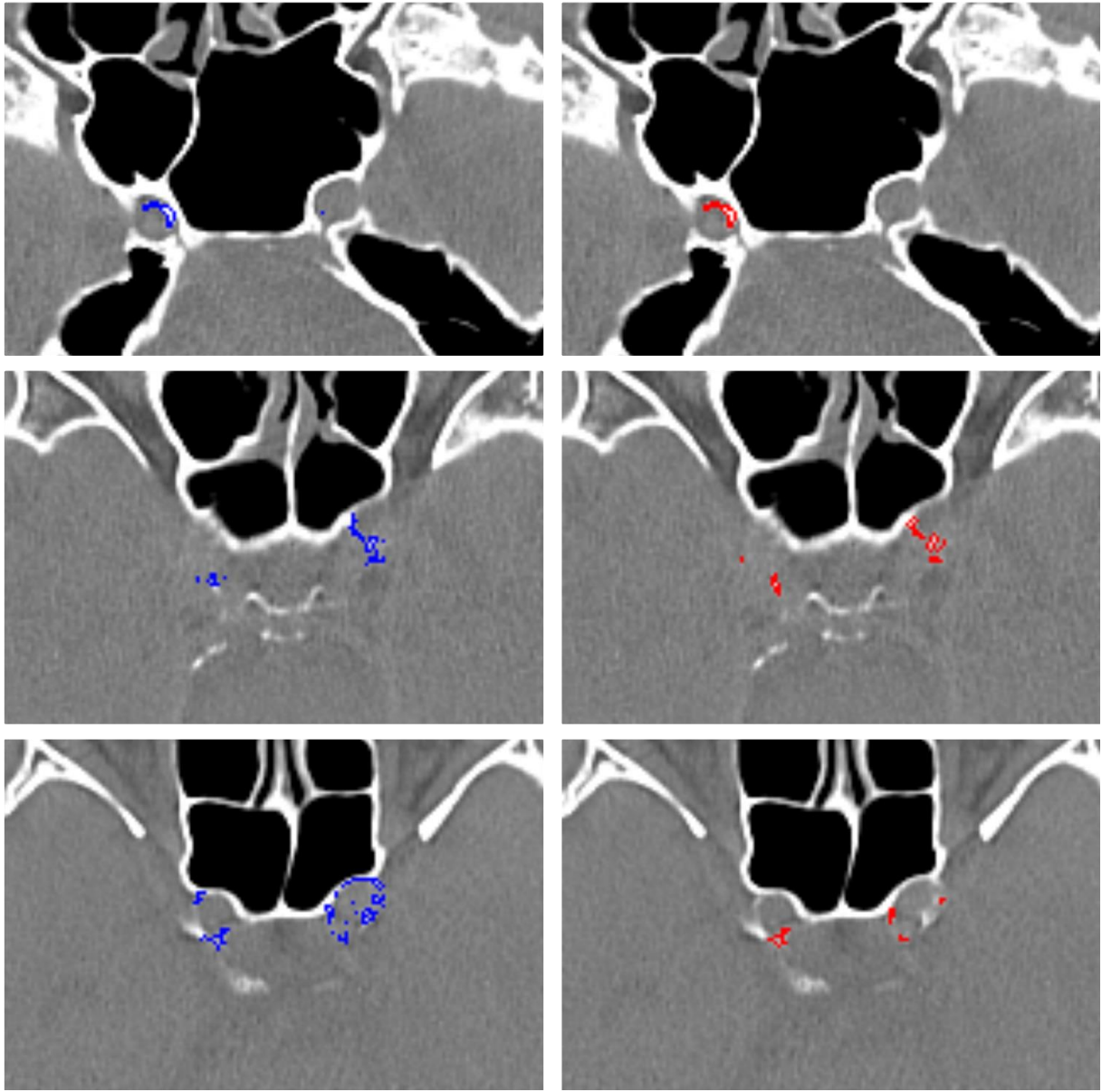

**Figure IV: Manual Segmentations and Automated Segmentations Computed With Pre-Smoothing Input Images**

Manual segmentations (left, blue) and automated segmentations (right, red) computed with smoothing images with Gaussian filter with sigma = 0.6 prior to applying the deep ensemble. The smoothed images resemble the Rotterdam Study images more (see Figure II).

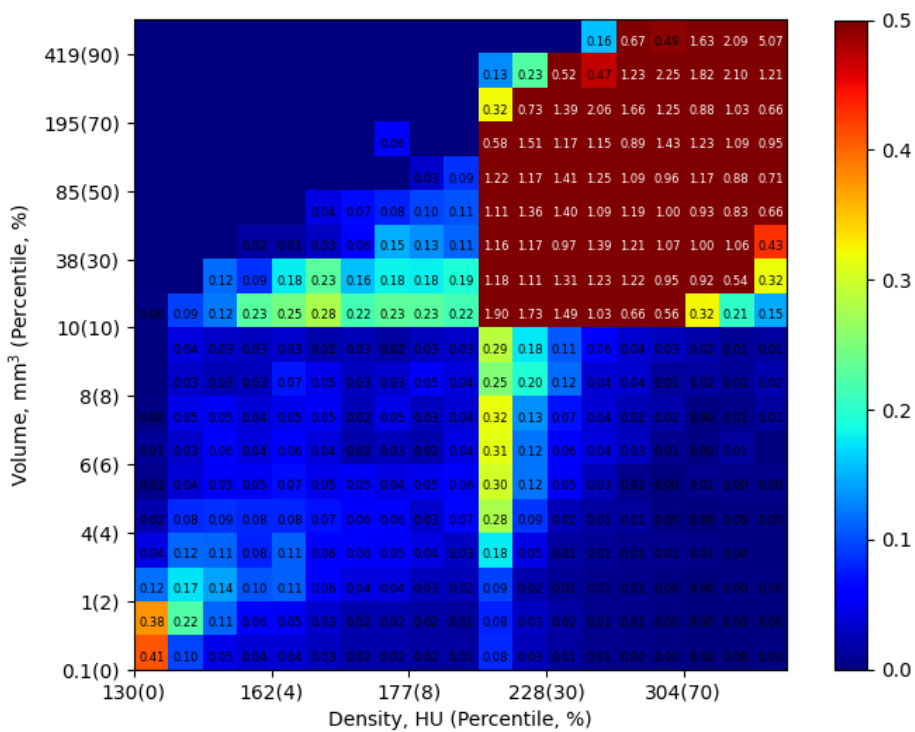

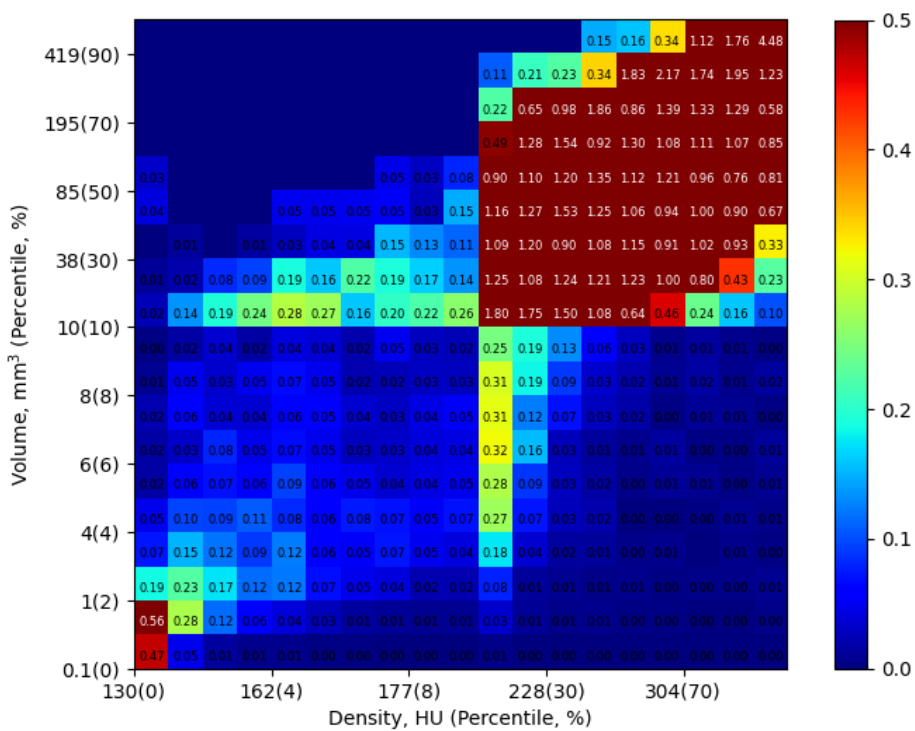

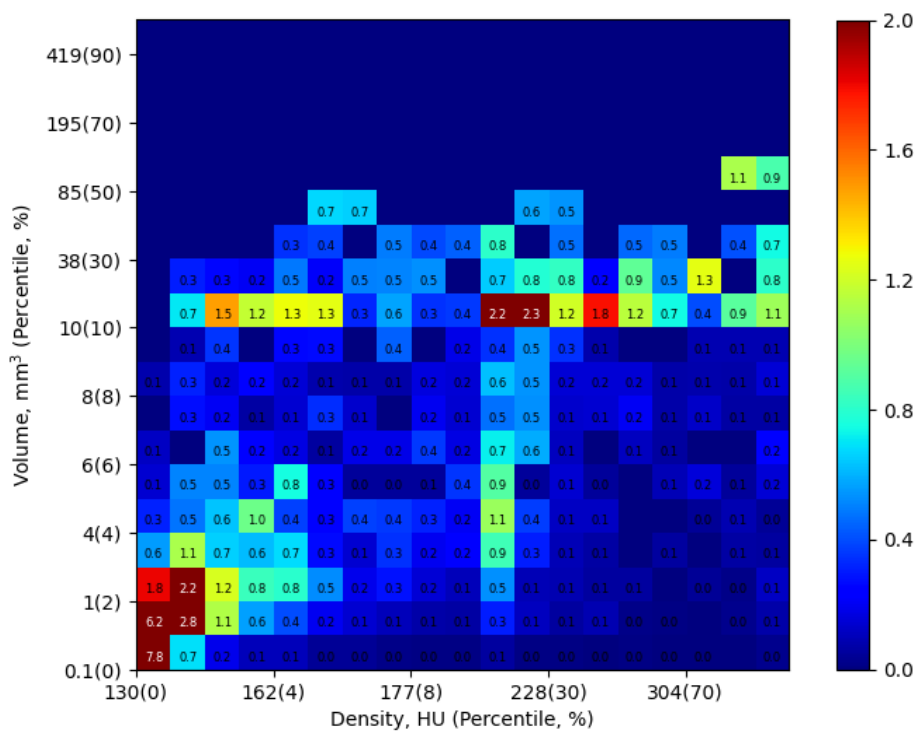
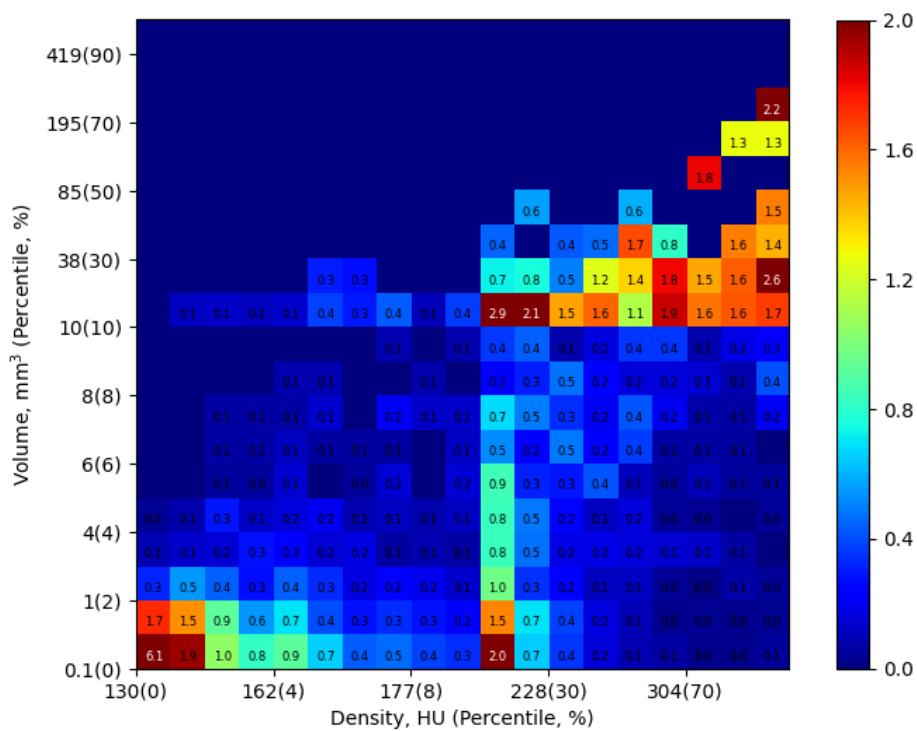

**Figure V: Distributions of Volume Versus Density of All Manually (A) and Automatically (B) Identified Lesions and the subsets of False Positive (C) and False Negative (D) Lesions of the Automated Segmentation Versus Manual Segmentation**

The bin ranges for volume and density, which can be seen on the x and y axes of the plot, were defined as percentiles of volumes and densities of all manually identified lesions, respectively, adjusted for volume (for example, lesions below 1 mm$^3$ — 2$^{nd}$ percentile of volume — constituted 2% of total manual ICAC volume in all participants). *Note that percentile scales are not linear and go from 0 to 10 and then 20 to 90; thus, left and bottom halves of these figures correspond to only 10% of ICAC volume, whereas their counterparts correspond to 90%.* Bin values are percentages of either the total volume of manually identified lesions or the total volume of false positive or false negative lesions. In the figures A and B, bin values are computed as (volume of lesions falling into a given range of volume and density have) / (total volume of manually identified lesions) * 100%. In the figures C and D, bin values are computed as (volume of lesions falling into a given range of volume and density have) / (total volume of all false positive or false negative lesions) * 100%.

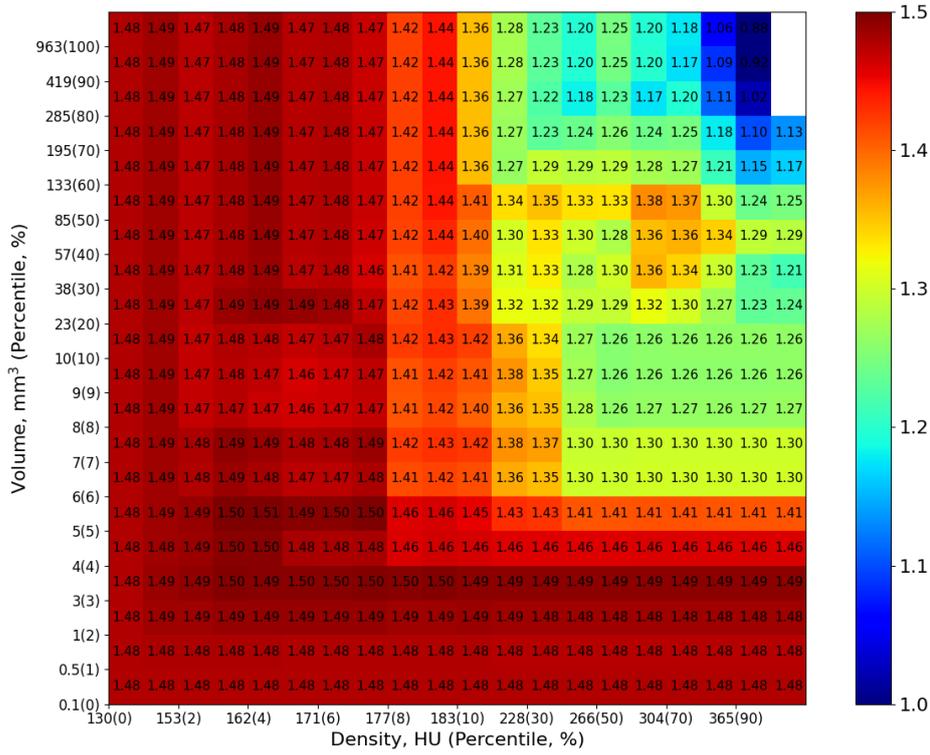
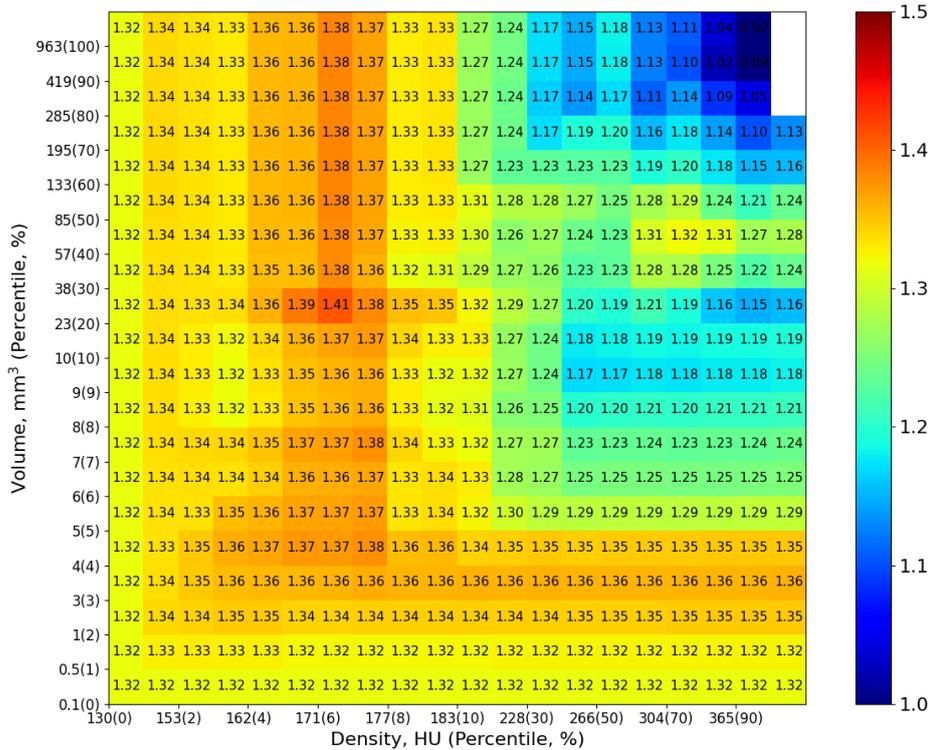

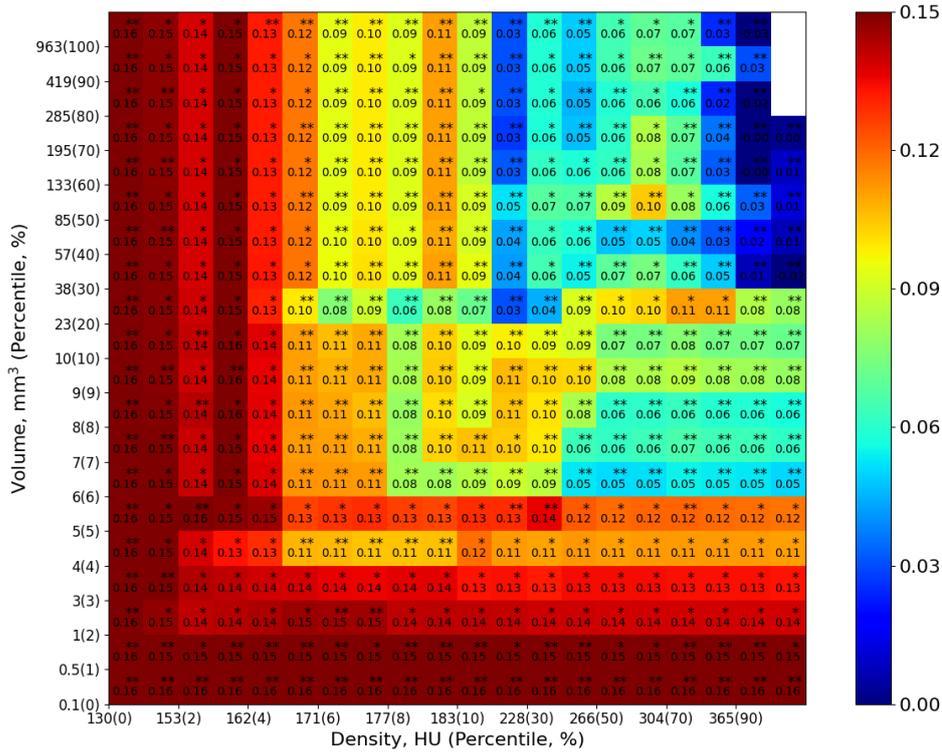

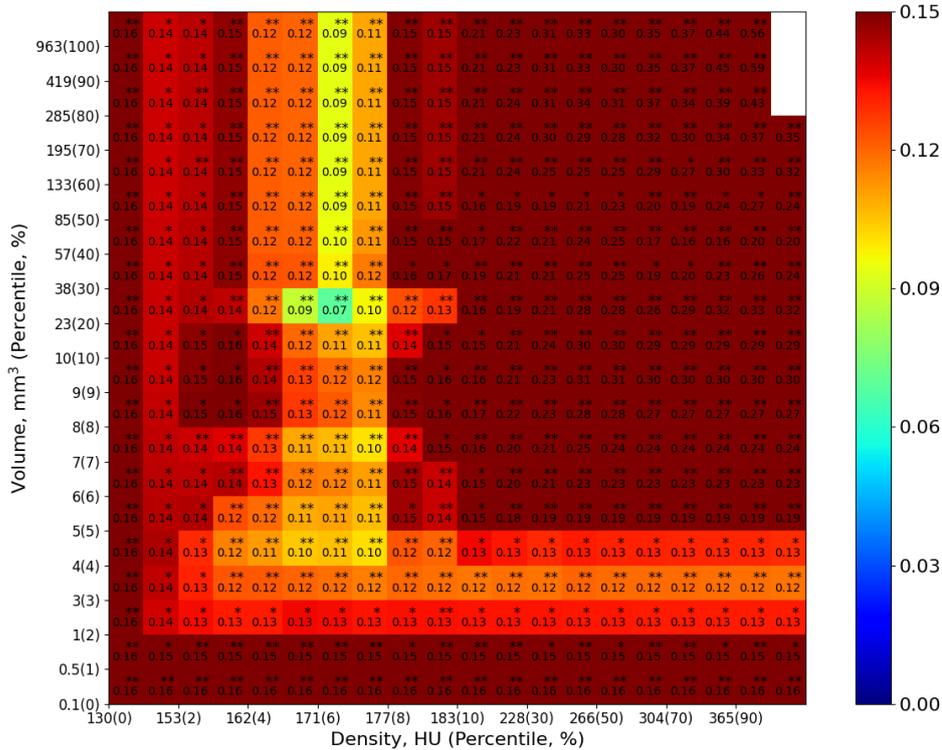

**Figure VI: Hazard Ratios for Manually and Automatically Computed ICAC Volumes Excluding Smaller and Lower-Density Lesions Using Varying Thresholds and the Difference Thereof**

The bin ranges for volume and density, which can be seen on x and y axes of the plot, were defined as percentiles of volumes and densities of all manually identified lesions, respectively, adjusted for volume (for example, lesions below 1 mm3 — 2nd percentile of volume — constituted 2% of total ICAC volume in all participants). *Note that percentile scales are not linear and go from 0 to 10 and then 20 to 100; thus, left and bottom halves of these figures correspond to only 10% of ICAC volume, whereas their counterparts correspond to 90%.* Values in these figures correspond to HRs or differences between HRs for manual and automatic volumes (follow-up until 1.01.2012) after excluding lesions below the minimums of ranges of volume and density that correspond to a given bin.

** P-value for HR difference below 0.05.

* P-value for HR difference equal or above 0.05 and below 0.1.

## A: HRs for Automatic Volumes with Larger and Higher-Density False Negative Lesions Added

## B: Differences Between HRs for Original Manual and Modified Automatic Volumes

**Figure VII: Hazard Ratios for Automatically Computed ICAC Volumes Including Larger and Higher-Density False Negative Lesions Using Varying Thresholds and the Difference Thereof**

The bin ranges for volume and density, which can be seen on x and y axes of the plot, were defined as percentiles of volumes and densities of all manually identified lesions, respectively, adjusted for volume (for example, lesions below 1 mm$^3$ — 2$^{nd}$ percentile of volume — constituted 2% of total ICAC volume in all participants). *Note that percentile scales are not linear and go from 0 to 10 and then 20 to 100; thus, left and bottom halves of these figures correspond to only 10% of ICAC volume, whereas their counterparts correspond to 90%.* Values in these figures correspond to HRs or differences between HRs for manual and automatic volumes (follow-up until 1.01.2012) after including lesions above the minimums of ranges of volume and density that correspond to a given bin.

\*\* P-value for HR difference below 0.05.

\* P-value for HR difference equal or above 0.05 and below 0.1.

# Supplemental Video Legends

**Video E1: Examples of Manual and Automatic Segmentations Provided to the Expert for Visual Comparison**

Visualizations of 20 randomly sampled regions exactly as they were presented to the expert observer for the analysis and the corresponding expert's assessments (on the right). The visualizations follow one another in the video (the region number can be seen on the left). Red and blue contours corresponded to either manual or automatic segmentations; which color represented which contour was random and not known to the observer. This video is supplied to demonstrate the visualization technique and provide examples of observer's assessments. The video can be viewed and downloaded [here](here).

**Video E2: Examples of Visual Assessments of Segmentations by the Expert Presented without Blinding Visualization**

Visualizations of 20 randomly sampled regions analyzed by the observer and the corresponding expert's assessments (on the right). These visualizations, with clearly indicated manual and automatic contours, were not shown to the observer and are presented here to show examples of manual and automatic segmentations and the expert's blinded judgements of their relative accuracy. The video can be viewed and downloaded [here](here).

# Supplemental Tables

**Table E1: Segmentation Performance of the Deep Ensemble and its Members**

|  | Dataset-wise statistics | | | Participant-wise statistics | | | | ICAC-free* participants (n = 422) |
|---|---|---|---|---|---|---|---|---|
|  | | | | Participants with ICAC* (n = 1,897) | | | | |
|  | F1, % | Recall, % | Precision, % | F1, % | Recall, % | Precision, % | FPV, mm$^3$ | FPV, mm$^3$ |
| Cross-entropy (CE)† | 84.6 | 83.1 | 86.1 | 75.7 ± 20.2 | 75.8 ± 24.0 | 81.9 ± 23.2 | 18.2 ± 31.0 | 8.9 ± 46.1 |
| Dice† | 83.6 | 82.0 | 85.2 | 77.8 ± 19.2 | **80.5 ± 20.2** | 80.6 ± 21.8 | 19.3 ± 27.1 | 9.6 ± 18.5 |
| Focal loss† | 84.4 | 82.8 | 86.1 | 76.0 ± 20.7 | 76.1 ± 22.8 | 82.2 ± 21.7 | 18.5 ± 29.8 | 7.3 ± 14.7 |
| Weighted CE† | 84.3 | **84.7** | 83.9 | 76.7 ± 20.0 | 80.0 ± 20.9 | 78.9 ± 22.5 | 21.8 ± 30.8 | 11.3 ± 26.4 |
| Ensemble | **85.9** | 83.8 | **88.0** | **80.0 ± 18.9** | 80.6 ± 20.8 | **84.3 ± 20.3** | **15.8 ± 24.5** | **6.2 ± 12.8** |

Note.—Participant-wise average metrics are mean ± SD. CE = cross-entropy loss function, FPV = false positive volume.

\* As indicated by the observer(s).

† The ensemble members are named after objective functions they use.

**Table E2: ICAC Segmentation Performance on Rotterdam Study and Trauma Datasets**

| Dataset and preprocessing parameters | | Dataset-wise statistics | | Participant-wise statistics | | | |
|---|---|---|---|---|---|---|---|
| | | All pixels in all scans | | Participants with ICAC* (n = 1,897) | | | ICAC-free* Participants (n = 422) |
| | N | Recall, % | Precision, % | Recall, % | FPV, mm$^3$ | Volume, mm$^3$ | FPV, mm$^3$ |
| **Rotterdam Study** | 2,319 | 83.8 | 88 | 80.6 ± 20.8 | 15.8 ± 24.5 | 150.3 ± 204.6 | 6.2 ± 12.8 |
| **Trauma, with smoothing**** | 100 | 73.0 | 80.9 | 66.7 ± 18.5 | 48.1 ± 51.5 | 281.3 ± 246.4 | 11.6 ± 12.2 |
| **Trauma** | 100 | 85.5 | 30.5 | 82.5 ± 18.6 | 846.7 ± 339.9 | 281.3 ± 246.4 | 804.2 ± 316.3 |

Note.—Participant-wise values are mean ± SD for the Rotterdam Study results and mean of means ± mean of SDs computed across predictions of ensembles trained on the ten folds of Rotterdam Study Data. 'Volume' is ICAC volume for participants with ICAC as assessed by the observers, reported for comparison with FPV. FPV = false positive volume.

* As indicated by the observer(s).

** Metrics were computed using ensemble predictions on scans smoothed using Gaussian filter with sigma = 0.6. The smoothing parameter sigma was tuned on a separate subset of 14 scans.